**Understanding the Relationship between Social Distancing Policies, Traffic Volume, Air Quality, and the Prevalence of COVID-19 Outcomes in Urban Neighborhoods**

- Daniel L. Mendoza, Department of Atmospheric Sciences, University of Utah, Salt Lake City, Utah USA (Corresponding Author Email: Daniel.Mendoza@utah.edu; 135 S 1460 E, Room 819, Salt Lake City, UT 84112-0102)

- Tabitha M. Benney, Department of Political Science, University of Utah, Salt Lake City, Utah USA

- Rajive Ganguli, Department of Mining Engineering, University of Utah, Salt Lake City, Utah USA

- Rambabu Pothina, Department of Mining Engineering, University of Utah, Salt Lake City, Utah USA

- Benjamin Krick, Department of Political Science, University of Utah, Salt Lake City, Utah USA

- Cheryl S. Pirozzi, Division of Pulmonary and Critical Care Medicine, Department of Internal Medicine, School of Medicine, University of Utah, Salt Lake City, Utah USA

- Erik T. Crosman, Department of Life, Earth, and Environmental Sciences, West Texas A&M University, Canyon, Texas USA

- Yue Zhang, Division of Epidemiology, Department of Internal Medicine, School of Medicine, University of Utah, Salt Lake City, Utah USA

**Highlights**

- Higher income zip codes had greater traffic reductions during Stay-at-Home period
- Lower traffic resulted in lower $PM_{2.5}$ and $NO_2$ pollution but increased at reopening
- Higher minority zip codes had up to 10 times the rate of COVID-19 positive cases
- Differential effectiveness of social distancing in limiting the spread of COVID-19
- Additional policies should be used to maximize protection of vulnerable populations

**Abstract**

**Background:** In response to the COVID-19 pandemic, governments have implemented policies to curb the spread of the novel virus. Little is known about how these policies impact various groups in society. This paper explores the relationship between social distancing policies, traffic volumes and air quality and how they impact various socioeconomic groups.

**Objective:** This study aims to understand how disparate communities respond to Stay-at-Home Orders and other social distancing policies to understand how human behavior in response to policy may play a part in the prevalence of COVID-19 positive cases.

**Methods:** We collected data on traffic density, air quality, socio-economic status, and positive cases rates of COVID-19 for each zip code of Salt Lake County, Utah (USA) between February 17 and June 12, 2020. We studied the impact of social distancing policies across three periods of policy implementation.

**Results:** We found that wealthier and whiter zip codes experienced a greater reduction in traffic and air pollution during the Stay-at-Home period. However, air quality did not necessarily follow traffic volumes in every case due to the complexity of interactions between emissions and meteorology. We also found a strong relationship between lower socioeconomic status and positive COVID-19 rates.

**Conclusion:** This study provides initial evidence for social distancing's effectiveness in limiting the spread of COVID-19, while providing insight into how socioeconomic status has compounded vulnerability during this crisis. Behavior restrictions disproportionately benefit whiter and wealthier communities both through protection from spread of COVID-19 and reduction in air pollution. Such findings may be further compounded by the impacts of air pollution, which likely exacerbate COVID-19 transmission and mortality rates. Researchers need to continue to understand the complex repercussions of behavioral regulation and policy makers to think of adapting social distancing policies to maximize equity in health protection.

**Introduction**

The social distancing policies enacted in the Spring of 2020, in response to the growing SARS-CoV-2 pandemic have saved many lives (Ferguson et al. 2020; McKibbin and Fernando 2020; Van Bavel et al. 2020). While social distancing policies are effective, little is understood about how the benefits of these policies accrue across various populations within society. In this study we examine the impact of one type of social distancing policy, Utah's "Stay Safe, Stay Home" (SSSH) directive, on traffic volumes and air quality, and how these factors may be helpful in explaining the prevalence of COVID-19 positive cases (hereafter "COVID-19 cases") in Salt Lake County (SLCo), Utah (USA). While government social distancing policies successfully mitigate the spread of COVID-19, variability in human behavior in reaction to these policies may help explain why some areas of SLCo experienced higher rates of COVID-19 than others. To understand the relationship between social distancing policies and human behavior variability, we compare traffic counts, pollutant observations, and COVID-19 cases by zip code during three phases of policy implementation between February 17 and June 12, 2020.

The air quality in Utah's urban centers, which are all located in topographical basins that trap air pollution, can vary significantly (McDuffie et al. 2019; Whiteman et al. 2014). SLCo is located at the intersection of four major highways (I-80, I-15, I-215, and U.S. 89) and traffic density and congestion has increased by ten percent or more annually in recent years (INRIX Research 2020). Therefore, transportation emissions are an important contributor to poor air quality. Recent studies provide increasing evidence that air pollution contributes to a range of illnesses including asthma, COPD, heart disease, pneumonia, depression, low birth weight, and increased mortality (Brauer 2010; DeVries et al. 2017; Hackmann and Sjöberg 2016; Liu et al. 2009; McCreanor et al. 2007; Pirozzi et al. 2018a; Pirozzi et al. 2018b). In 2015, air pollution contributed to 7.6 percent of total global mortality, making it the fifth ranked global health risk factor (Zanobetti and Schwartz 2009). Since traffic both contributes to and compounds the health impacts associated with air quality, this relationship is of particular interest to researchers and policy makers alike.

Economically disadvantaged populations face the highest levels of exposure to air pollution, and the resulting health risks, which has been attributed to three main factors: income, education, and occupation (Bekkar et al. 2020). Pollutant exposure rates in the US are more likely determined by socioeconomic status than any other factor (Bekkar et al. 2020; Bell and Ebisu 2012; Clark et al. 2014). The uneven exposure to environmental risk resulting from social, economic, and political processes that worsen economic and social inequality is defined as "environmental justice". For example, a recent U.S. study found that pregnant women exposed to high temperatures or air pollution are more likely to have children who are premature, underweight, or stillborn, and the effects hurt African-American mothers and children the most (Bekkar et al. 2020). Additional research has found disparities associated with air quality and socioeconomic status, language minority status, immigration status, race, and ethnicity on school absences and other educational outcomes (García and Weiss 2018). Emerging research suggests that patients with severe COVID-19 are twice as likely to have had pre-existing respiratory diseases and three times as likely to have had cardiovascular problems (Yang et al. 2020) severely compounding concerns over this relationship.

Despite these recent findings, research on how SARS-CoV-2 virus containment affects the prevalence of disease across community subgroups is only now beginning to emerge. During the outbreak of SARS in the early 2000s, a dramatic decrease in asthma incidence and acute respiratory infections was noted as a result of the type of community containment and hygiene adopted by the government and people in the countries where the studies took place (Clay et al. 2019). Another study of the 1918 Spanish Flu pandemic noted that cities with higher pollution had higher incidences of viral respiratory infections (Shaker et al. 2020). Given that the community containment measures being implemented worldwide are leading to a decrease in air pollution and traffic density, which may interact with disease transmission and mortality, it is also appropriate to test how these factors interact with COVID-19.

To understand how social distancing policies like SSSH accrue across urban populations, we examine the impact of these policies on traffic density and zip code level air quality and compare these outcomes by zip code to see if all neighborhoods are impacted equally, or if meaningful differences occur. Next, we consider how air quality may impact COVID-19 outcomes. To understand this complex relationship, we compare air quality, traffic levels, income, and minority status with COVID-19 cases by zip code to further identify the effects of COVID-19 on these variables and additional subgroups. By exploring these compounding factors at the zip code level, we aim to illustrate how variation in human behavior resulting from social distancing policies facilitated epidemiological transmission patterns that may have impacted some communities more than others. Such findings may help explain how social demographics and traffic volume, used as a proxy for human activity, play a role in the prevalence of COVID-19 in some communities compared to others.

**Methods**

Our research explores three key propositions about the relationship between SSSH policies, traffic volumes, and air quality and how these factors may help explain the prevalence of COVID-19. The first proposition explores the impact of social distancing policies on traffic and air quality through three distinct periods of policy implementation. The study goal is to understand the relationship between social distancing policies and traffic density at sites of interest throughout the city and at the zip code level. Our second proposition explores if vulnerable (e.g. low income and minority status) communities in SLCo have equal traffic variability and air quality exposure as other high income, low minority status communities. Our final proposition explores whether traffic and air pollution levels are associated with COVID-19 cases and what this relationship might imply for policy and future research.

*Study Location*

This study takes place in SLCo, Utah (USA), the largest urban area in the state with a population over 1.16 million inhabitants. SLCo has a large immigrant population with varying income levels and areas of both high and low geographic elevation, with lower elevations generally observing higher pollution levels. Most importantly, SLCo has a state-of-the-art research grade air pollution and traffic count sensor networks to study human impacts on natural settings at varying scales (Lin et al. 2018; Mendoza et al. 2019).

*Study timeline*

To better understand the role of human behavior in this research, our study is focused on the three phases of implementation that occurred with the COVID-19 social distancing policies established at the start of the COVID-19 pandemic in 2020. As illustrated in Table 1, we define our research utilizing three policy implementation phases. We also lag the COVID-19 case count data by 14 days (until June 12, 2020) to account for the broadly accepted incubation stage of the virus.

**Table 1. List of Project Study Periods**

| **Period** | **Dates** | **Description** |
|---|---|---|
| Pre-SSSH | February 17 – March 15 | The period prior to the SSSH policy implementation |
| SSSH | March 16 – April 26 | SSSH policy implementation period |
| Post-SSSH | May 1 – May 28 | The easing period following SSSH policy implementation |

*Data Sources*

SLCo is composed of 38 zip codes with population greater than 0. Of the 38 zip codes, 34 were utilized for this study and 4 were excluded for the reasons noted below. The sparsely populated zip code 84006 (Pop: 1,041) did not have a confirmed COVID-19 case throughout the study period and was removed from the study. Zip code 84009 was created in 2015 and there is no current sociodemographic data, therefore it was also excluded from the study. Two zip codes (84112 and 84113) belong to the University of Utah and reflect the student population in dormitories. Since the University moved to a completely online format on March 11[th] to prevent students from returning to campus following Spring Break, these zip codes were not reflective of the population in that area and were also removed from the study.

Sociodemographic Data: Sociodemographic information for study zip codes was retrieved from the SLCo's "Healthy Salt Lake" program dataset for 2020 (Healthy Salt Lake 2020). The study variables used are population, percent white population, average household size, and median household income. Per capita income was derived by dividing median household income by average household size. As shown in Supplementary Information, Table S1, the per capita income by zip code in SLCo varies from $14,534 to $43,068 and percent white population ranges from 50.13% to 92.55%. To minimize noise in our analysis, the zip codes were grouped into 5 groups of 7 zip codes (with the middle group having only 6 members) by income: Group 1 ($14,534-$20,443), Group 2 ($20,502-$24,554), Group 3 ($25,396-$29,558), Group 4 ($30,528-$36,325), and Group 5 ($36,443-$43,068) and similarly by percent white population: Group 1 (50.13%-64.09%), Group 2 (74.48%-77.67%), Group 3 (77.91%-83.93%), Group 4 (85.41%-89.19%), and Group 5 (89.74%-92.55%)

Traffic Count Data: Traffic count data was accessed through two data portals. The traffic counts for Utah roads and highways is available through Automated Traffic System Performance Measures (ATSPM) (Utah Department of Transportation 2020a), and the counts for interstate traffic is obtained from the Performance Measurement System (PeMS) (Utah Department of Transportation 2020b). We classified roads in SLCo into two groups: residential (reflecting local traffic) and non-residential (United States Department of Transportation 2012). In some residential areas with no available street data, a small state highway was used to represent residential traffic. The non-residential category includes interstates and state highways.

We compiled data for residential and non-residential traffic volume counts for each zip code from February 17 to May 28, 2020. This time range accounts for the variation in human behavior due to the three stages of the SSSH policy implementation (e.g. pre-policy, policy fully implemented, policy easing). For each traffic sensor location, we recorded the count of cars traveling past the sensor per hour. For each zip code, we identified the residential and a non-residential traffic sensor closest to the centroid to retrieve traffic volume. We used traffic counting sites selected due to their proximity to either an air quality sensor (n=10) or a point of interest (n=12) including hospitals, large stores, downtown Salt Lake City, and the airport.

Air Quality Monitoring Station Level Air Pollution Exposure Data: Air quality data from ten air quality sites was retrieved during the study period and consisted of the hourly concentration (parts per million) of carbon monoxide (CO), nitrogen-dioxide ($NO_2$), ozone, and fine particulate matter ($PM_{2.5}$). The closest traffic sensor with available data in relation to the air quality (AQ) site is used. For some sites, the traffic counts can be recorded for the exact location of the AQ site, while for other AQ sites, the traffic count may be up to two city blocks away.

Zip Code Level Air Pollutant exposure data: We use data from the U.S. Environmental Protection Agency's AirNow database (United States Environmental Protection Agency 2020), the MesoWest air quality network (Horel et al. 2002), and mobile air quality sensors mounted on light rail trains (Mendoza et al. 2019) to estimate hourly pollutant concentrations for $PM_{2.5}$ and ozone at each zip code. Median values were also computed for each of the three study periods for each zip code.

COVID-19 Confirmed Cases Data: The SLCo Health Department's data dashboard provides a count of confirmed cases of COVID-19 (Salt Lake County Health Department 2020). This data is aggregated by zip code, and daily counts were extracted from March 4, 2020 (when the first reported COVID-19 case was reported in Utah) to June 11, 2020 (two weeks following the project end date to account for the viral incubation period).

Statistical Methods: All data were first averaged by day. Change in COVID-19 rates is derived using means to avoid skewing the data. Due to the varied lengths of the pre-SSSH, SSSH, and post-SSSH periods, and the non-normality of the data, medians were the preferred statistic for comparison, and it is also ideal for handling data gaps and outliers. Data processing and analysis was done using Matlab Version R2019b (Mathworks 2020) and R Version 3.6.3 (R Core Team 2019) software.

**Results**

Based on our analysis, the three main propositions in this research present a highly nuanced view of human behavior from social distancing at the neighborhood level. At this level of precision, several important patterns emerge that shed light on the unequal benefit this policy tends to produce and some important disjunctions between traffic and air quality patterns. The following section reviews each proposition in detail.

*Proposition 1: Impacts of social distancing policies on traffic density and air quality*

Our first research question explored the impact of COVID-19 related social distancing policies on traffic and air quality through the three periods of policy implementation.

Traffic Impacts: We found that traffic volume in SLCo decreased by 30-40% with the onset of social distancing policies (difference between pre-SSSH and SSSH, Figure 1) and bounced back by 20-30% as these policies were slowly relaxed (post-SSSH period). As illustrated in Figure 1 below, when all major categories of industry and residential actors are compared, they produce similar outcomes across the study period. However, when we move from the aggregate level to the zip code level, a much more disparate pattern emerges.

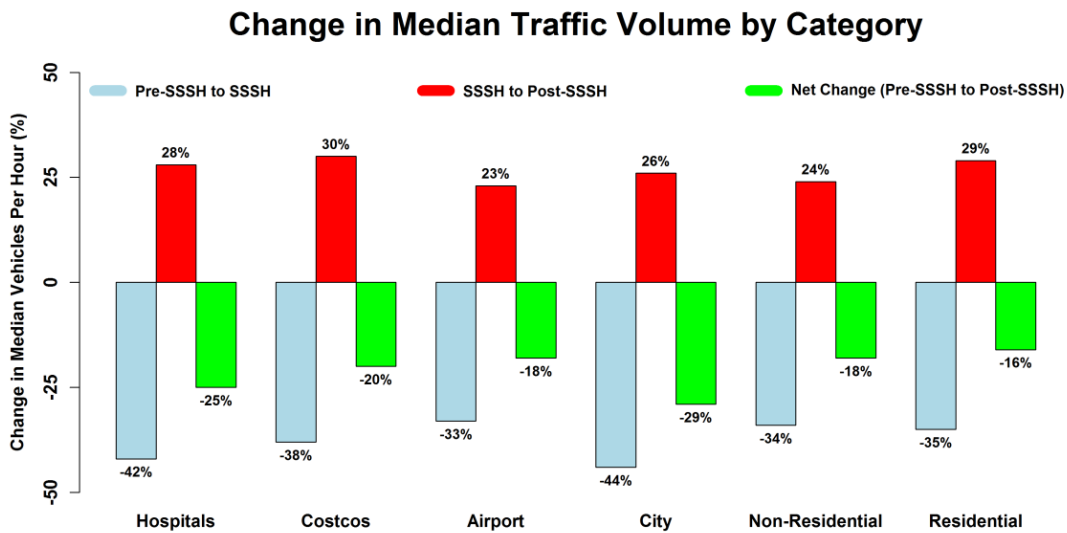

Figure 1. Percent change in traffic volume by category for pre-SSSH, SSSH, and post-SSSH policy implementation phases.

At the zip code level, we find that the pre-SSSH median level of zip code residential weekday traffic ranged between 55 to 1834 vehicles per hour (vph) as shown in Supplementary Information Table S1. In stage two of the implementation period, when the county was under the SSSH directive, the median ranged from 42 to 1168 vph. This suggests that some zip codes responded to the SSSH directive at a greater magnitude than others. In the post-social distancing period, the median traffic volume ranged from 54 to 1476 vph. Rapid unemployment emerging locally, which has disproportionately impacted middle- and low-income families, may be an explanatory factor in the differences observed in the post-SSSH perios. Most importantly, the increased variability found in the second and third period of policy implementation suggests that SSSH directives may have had limited impacts in some areas in comparison to others due to the potentially larger fraction of essential workers residing in lower income, higher minority zip codes.

Air Quality (AQ) Impacts: Figure 2 shows the relationship between traffic and AQ measurements at each paired location. The traffic volume shows expected patterns, with a marked decrease during the transition from pre-SSSH to SSSH followed by an increase in the post-SSSH phase. $PM_{2.5}$ closely mirrored traffic volume reflecting the decrease and subsequent increase in vehicular traffic and other economic activity through the three study periods. CO decreased steadily across the study period, but this is expected given that CO is essentially a direct measure of on-road emissions, which will be increasingly mixed and diluted into the surrounding atmosphere by the stronger solar forcing observed in the later study periods. $NO_2$ showed a large initial decrease between the Pre-SSSH and SSSH, which illustrates why $NO_2$ is a common pollutant for analyzing on-road vehicle tailpipe emissions. However, a smaller rebound during the SSSH to post-SSSH transition is noted compared to the corresponding increase in traffic, again believed to be due increased atmospheric mixing from SSSH to Post-SSSH. Ozone climbed continuously between the periods. Ozone production is a photochemical reaction and background ozone levels across the Western US generally increase from winter into spring as solar irradiance increases, irrespective of local emissions (Cynthia Lin et al. 2000; Vingarzan 2004). The decreased onroad tailpipe emission of $NO_2$ in the SSSH period likely reduced the $NO_2$ consumption of ozone, allowing a more rapid increase in ozone during SSSH to post-SSSH period than would have otherwise been expected.

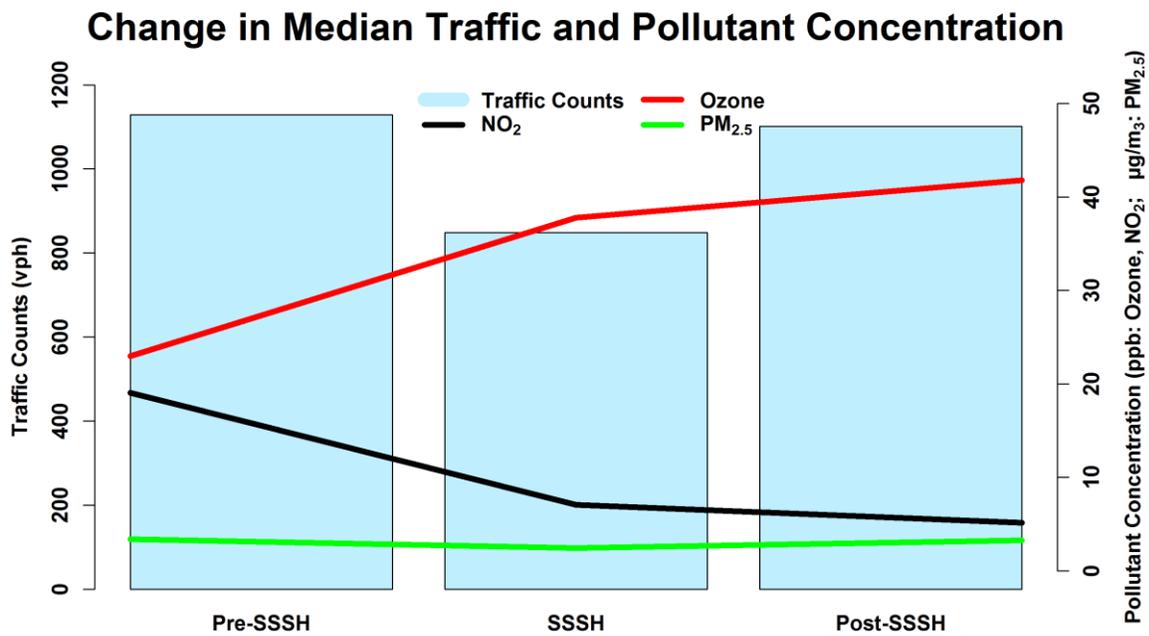

Figure 2. Traffic and pollutant level trends at co-located sensor sites (all study periods)

*Proposition 2: Impacts of Social Distancing Policies on Vulnerable Populations*

Our second research question explores the uniformity of the policy impact across urban populations. Since SLCo has a high variance of per capita income ($14,534 - $43,068) across zip codes (Supplementary Information, Table S1), our goal was to explore if vulnerable (e.g. low income or high minority status) communities have equal traffic and air quality impacts as high income, low minority status communities. We reasoned that individuals in lower income brackets are, on average, more likely to work in sectors that are classified as essential worker, while individuals in higher income brackets are more likely to work in sectors and occupations with increased ability to work from home (Von Gaudecker et al. 2020). Thus, workers who are able to stay home will be responsible for less traffic and air pollution in their zip code.

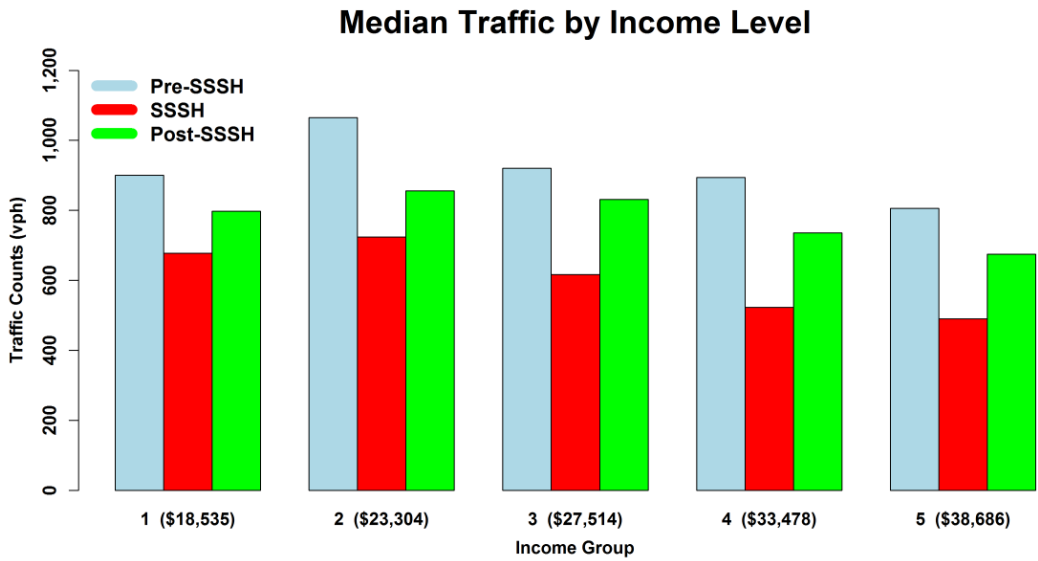

A

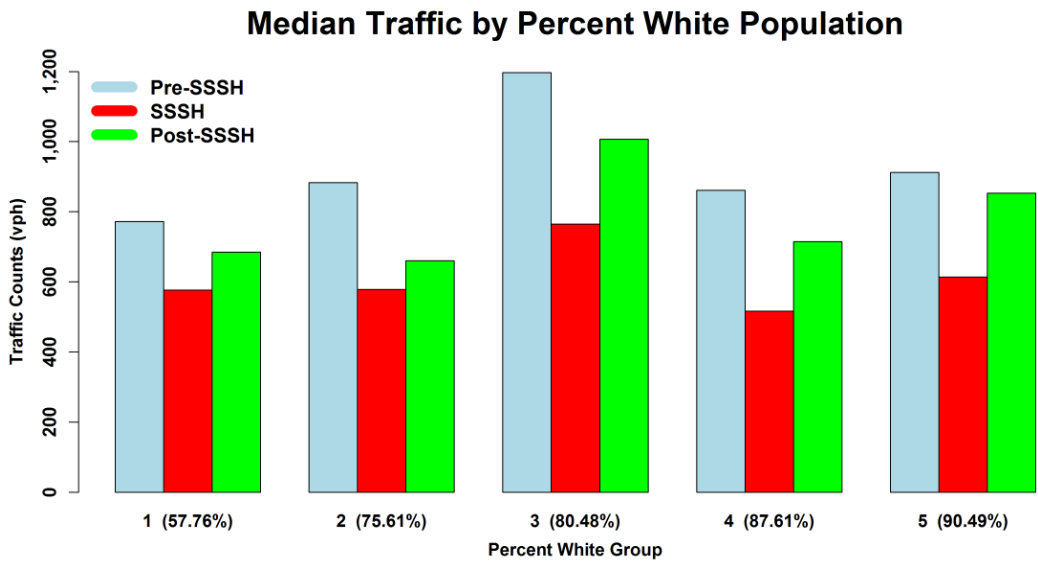

B

Figure 3. Change in traffic volume across each policy stage for each (A) per capita income and (B) percent white group.

Figure 3A shows the relationship between changes in traffic with phases grouped by income. In the figure, the blue boxes show the in median (in vph) across all three phases. There are several key findings noted when traffic and income levels by zip code are compared throughout the three study periods as discussed below:

In the pre-SSSH policy implementation stage, the median traffic levels averaged 900, 1065, 920, 894, and 806 vph in Groups 1-5, respectively (Supplementary Information, Table S1). Thus, the level of traffic was comparable in all four groups, but the highest income group (Group 5) notably averaged about 15% less traffic than the others - even prior to the onset of this study.

Through the course of the policy, we found the reduction in traffic to be least (-25%) in the lowest income group (Group 1) and most (-42% and -41%) in the two highest-income groups (Groups 4 and 5). Additionally, the reduction in traffic from pre-SSSH to SSSH is closely patterned on income with higher income groups showing greater reductions. Likewise, upon easing the policy, the traffic volume was once again related to income. In the transition from the SSSH to post-SSSH easing period, further key findings emerged. The two lowest income groups have the lowest bounce back in traffic (18% and 21% for Groups 1 and 2 respectively), while the two highest income groups have a markedly higher bounce back in traffic levels (40% and 34% for Groups 4 and 5, respectively).

Figure 3B. shows a similar relationship between traffic volume and percent white by group as was presented in Figure 3A for traffic volume and income. Based on these traffic volume results, we can conclude that lower income and minority status groups were not equally benefitted by the SSSH policy.

Through Figure 4, which shows the relationship between air quality and income (Figure 4A) and air quality and minority status (Figure 4B), we find nearly identical outcomes that further confirm these findings. Thus, we are able to add to the literature on environment inequality by illustrating the strong relationship between race, income, traffic congestion, and air pollution. Such findings will need further research to better understand the underlying mechanisms and policy solutions that will be address the regressive nature of some social distancing policies.

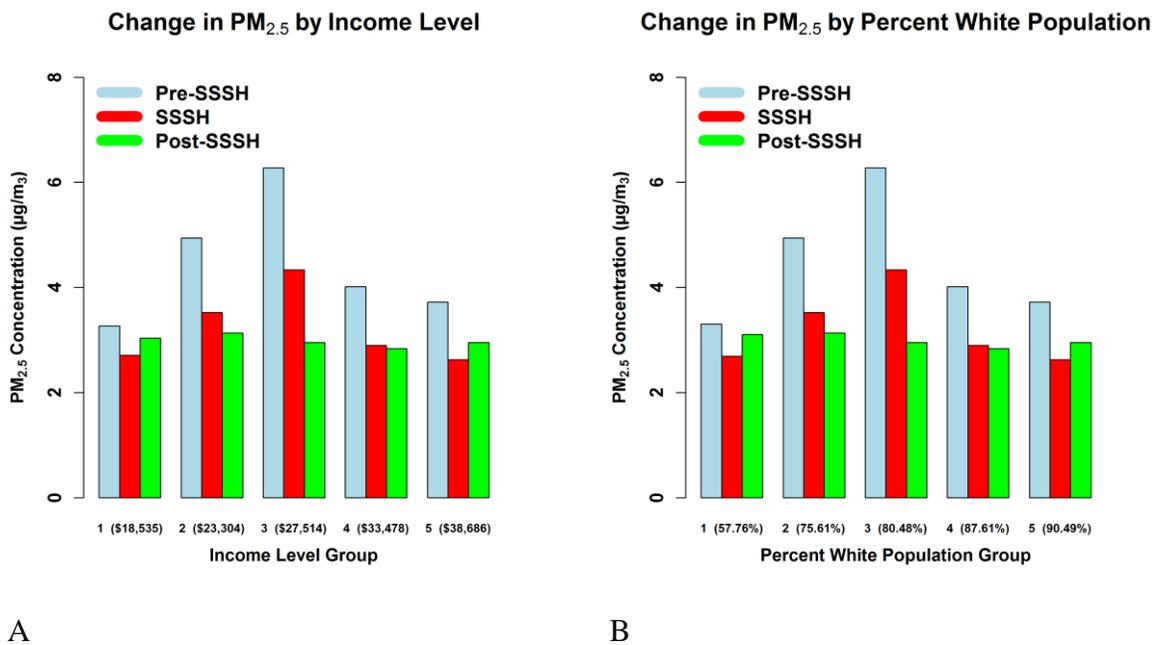

A                                                              B

Figure 4. Change in air quality across each policy stage for each (A) per capita income and (B) percent white group.

*Proposition 3: What is the Relationship Between Traffic Volumes, Air Pollution Levels, and COVID-19 Cases for Various Populations?*

While it is difficult to determine causality between the impacts of air pollution on COVID-19 cases in the short term, further insights into the relationship between these related factors are explored here. Perhaps the greatest concern in this area is how pollution exposure is commonly linked to income level and what the short- and long-term implications of this might be. Since a strong association has been established between income level, minority status levels, traffic density, and pollution exposure (see discussion above), we compare these factors with COVID-19 cases (Figure 5 and Supplementary Information Figure S1) by zip code.

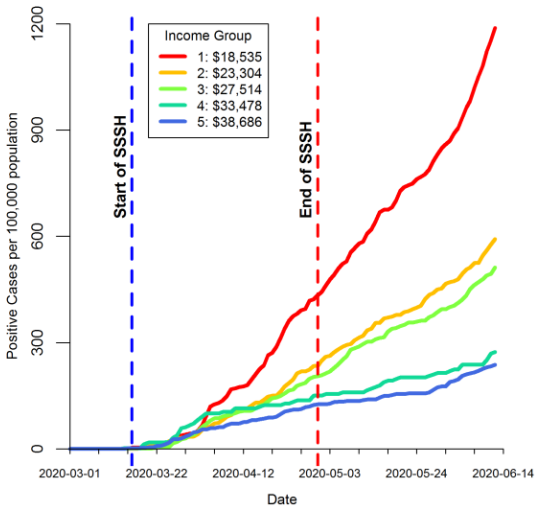 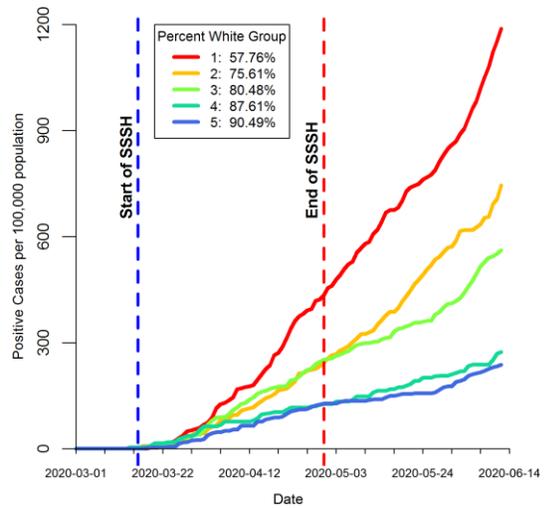

A B

Figure 5. COVID-19 cases for each (A) per capita income and (B) percent white group. The dashed vertical lines show the start (blue) and end (red) of SSSH directives. The color scale ranges from red (lowest income) to blue (highest income).

Based on this analysis, we find that overwhelmingly, lower income (Figure 5A) and lower percent white population (Figure 5B) zip codes had higher COVID-19 cases. The lowest income zip code (84104: $14,533) had the lowest percent white population (50.13%) and had the highest rate (1214.64 positive cases per 100,000), while the highest income zip code (84108: $43,068) had the thirteenth highest percent white population (85.77%) and second lowest rate (138.19 positive cases per 100,000). Additionally, lower income (Figure 6A) and lower percent white population (Figure 6B) zip codes show higher growth in positive COVID-19 rates throughout all phases of this study.

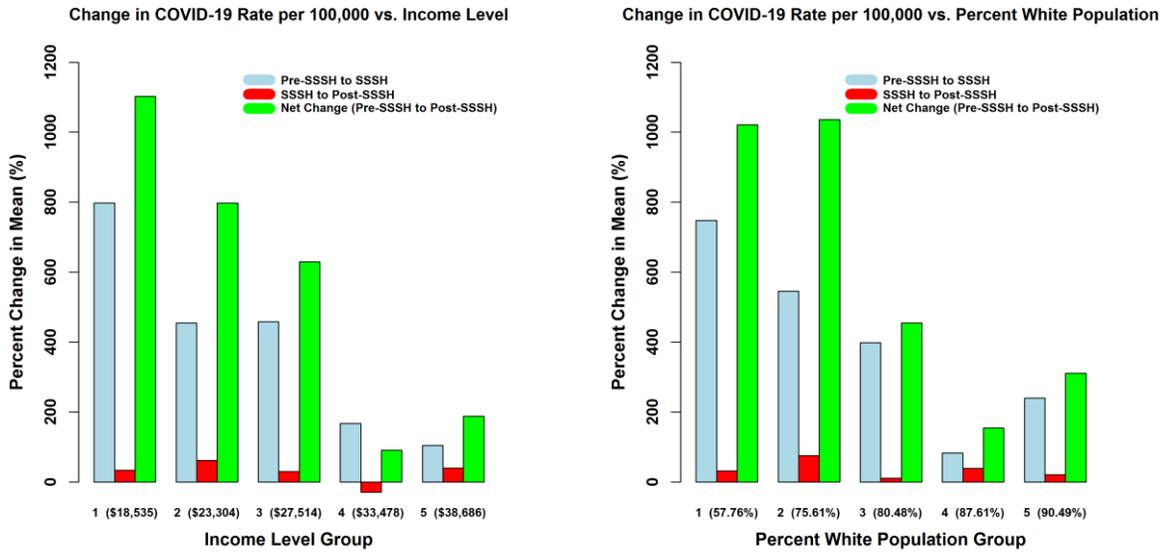

Figure 6. Change in COVID-19 cases between pre-SSSH, SSSH, and post-SSSH phases for each (A) per capita income and (B) percent white group.

Based on these findings, we conclude that the SSSH policy had a low, if not negligible, impact on low income and high minority status communities, but a high impact on higher income, low minority status zip codes in the policy implementation stage. However, upon easing of the policy, higher income, low minority status communities showed a rapid resurgence to activity, while lower income, high minority status communities, which had not seen a decline in activity across the policy implementation stage, showed a smaller increase in traffic. We attribute this to the rapid unemployment emerging locally, which has disproportionately impacted middle- and low-income families.

**Discussion**

Our preliminary survey of data on traffic density, air quality, socio-economic status, and positive cases rates of COVID-19 in SLCo, Utah leads to several relevant findings. Based on our analysis, we found that air quality did not necessarily follow traffic volumes in every case due to the complexity of interactions between emissions and meteorology. While some emissions behaved in a more intuitive manner (e.g. $PM_{2.5}$ dipped and recovered to original levels reflecting human activity), ozone increased naturally over the study period (as expected with increasing Spring temperatures). $NO_2$ sharply decreased due to a sudden decrease in vehicle emission sources associated with SSSH directives, but then increased imperceptibly across the study periods, likely due to interactions between atmospheric mixing and ozone chemistry as the period progressed. Such findings are relevant for understanding the complexity of air pollution temporal analysis across multiple months and how we should expect urban air quality to rebound following periods of social lockdown.

We also explored the impact of the SSSH policy on the community at the zip code level. Here we found that some zip codes responded to the Stay-at-Home directive at a greater magnitude than others (in terms of both traffic counts and air quality) and this change persisted across all three periods of policy implementation. When this analysis was extended to include income and race, an even starker pattern was revealed. For example, we find that when traffic or air quality levels are compared with income level or race at the zip code level, traffic and air pollution declined for all income and race categories under SSSH, with the lowest income or lowest percent white categories showing the smallest decrease. We find the reverse pattern in the post-SSSH easing period, suggesting the highest income and percent white category groups generally rebounding less as the policy was eased. Thus, the increased variability found in the second and third period of policy implementation suggests that SSSH directives may have had limited impacts on the most vulnerable groups. Perhaps the most important finding of this research is the clear relationship between traffic volume, race, income, and COVID-19 outcomes. While further research is needed to understand this relationship further, findings such as this should be used in the near-term to better inform and shape policy outcomes under a pandemic.

Further general findings are also apparent from this research. First, we found the impact between income and race to be almost identical in this case. Hence, income or race may serve as important proxies for vulnerable populations in future research of this type. Second, our findings shed further light on the unequal impact of air pollution and disease on vulnerable populations in society. Third, we found the zip code level analysis was critical in exposing these patterns. Due to the underlying factors in these data, when aggregated, critical policy impacts might otherwise go unidentified, thus further harming unfairly impacted groups within society.

While the impact of the pandemic is still unraveling in the United States, these preliminary findings may help to inform future research in this area. Despite this, some limitations in the data are possible at this early stage of the pandemic. COVID-19 cases may be under-reported or geographically biased due to differing testing rates. Also, a disproportionate number of deaths from SARS-CoV-2 are individuals in long-term care facilities and these factors may be found to be more consequential in future studies. However, since income and race were both found produce patterns that mirror each other almost identically with the prevalence of disease, it seems less likely that both variables might impact external validity in a similar manner to produce these outcomes. By exploring these compounding factors at the zip code level, we have illustrated how variation in human behavior resulting from social distancing policies encouraged epidemiological transmission patterns that impacted some communities more than others. Such findings help clarify how behavioral factors may play an important role in explaining the prevalence of COVID-19 in some communities over others. Findings of this study will advance research and policy in this area in at least four important ways. First, research on social distancing may help to advance the field of public health and related directives. Second, this research could help advance our understanding of the relationship between the environment and human health. Third, further study in this area may help inform our understanding of pollution rebounds and how air quality may interact with social factors to produce epiphenomena. Finally, efforts in this area will help advance policy design and implementation approaches for future pandemic and crisis planning.

*Limitations of this Study*

Due to the complexity of the system under analysis, research bias could be introduced over the course of the normal research process. Industrial data, like traffic or air quality sensors, is notoriously problematic and is conducive to bias and error because data quality can vary. For example, traffic count sensors may be geographically misaligned with the zip code centroid or the air quality sensor site. Likewise, both EPA and research grade air pollution sensors can go offline disrupting the ability of a researcher to use the most desirable sensor or resulting in missing data. Air pollution sensors can also occasionally observe errors and biases due to environmental factors such as dust events, wildfires, and atmospheric inversions. Errors from research-grade sensors are typically less than 2%, although slightly larger biases are possible due to a lack of data availability for some sensors on specific days. Additionally, the definition of residential versus non-residential traffic is not exact, and therefore discretion on these definitions was made by the researchers in each zip code. This required familiarity with the zip codes in question and an understanding of the activities and centrality of specific roads. Finally, error and bias can be introduced by the process of replacing missing data or through the selection of thresholds by the researchers.

*Implications for Future Research*

This research has at least three important implications:

COVID-19 and Air Quality: If air quality is compounding health risks for essential workers, now is more critical than ever to press for clean air regulation. Despite this, efforts at the federal level have been relaxed or rolled back and this may be exponentially harmful to all Americans due to the possible relationship between COVID-19 impacts and air pollution exposure.

Vulnerable Populations: Communal populations (e.g. populations experiencing homelessness, refugees, multi-generational families, etc.) are especially vulnerable because they are more likely to interact with individuals in the lower income groups, work in high exposure occupations, or live with others who share social support circles with other high risk populations, and may lack appropriate information, health care, or have an existing health deficit due to other compounding factors (e.g. language, culture, or socioeconomic barriers). As work-related spread continues throughout the pandemic, understanding impacts to workers may be especially important.

Policy Implications: Based on our findings, Shelter-in-Place and Stay-at-Home policies were not equally effective for all populations. Additional policies should be considered and used in tandem with such directives to offset the regressive nature of such policies. Inclusion of confounding factors into social distancing policies could also improve the overall policy performance, improve public health outcomes, ensure the long-term sustainability of such policies, reduce pandemic related risks for all citizens, and better protect all people equally in society.

While government policies to prevent the spread of COVID-19 led to highly successful social distancing programs, variability in human behavior in reaction to these policies help explain why some areas of SLCo experienced higher rates of COVID-19 than others. To understand this relationship, we compared traffic counts, pollutant observations, and COVID-19 cases by zip code during three key inflection points representing various stages of policy implementation during the 2020 pandemic. We examined the impact of these policies on traffic density and neighborhood level air quality and found that traffic density decreases were large and were followed by an equally notable rebound. These changes were directly related to income groups with wealthier (or higher percentage white population) zip codes showing a greater decrease in traffic as well as air pollution during SHSS than less wealthy (or lower percentage white population) zip codes. Such findings help clarify how social demographics and behavioral factors may play an important role in explaining the prevalence of COVID-19 in some communities and subgroups over others.

**Acknowledgements**

Seed funding for this project was provided by the *Emerging COVID-19/SARS-CoV-2 Research Program* of the Health Science Research Unit, University of Utah

**Supplemental Material**

Table S1: Median hourly traffic counts and per capita income by zip codes

| ZIP code | Income Group | Per Capita Income (Low to High) | Racial Composition Group | Percent White Population | Pre-SSSH Vehicles/ Hour (VPH) | SSSH Vehicles/ Hour (VPH) | Post-SSSH Vehicles/ Hour (VPH) |
|---|---|---|---|---|---|---|---|
| 84104 | 1 | $14,534 | 1 | 50.13 | 1200 | 1020 | 1177 |
| 84116 | 1 | $16,302 | 1 | 50.65 | 1462 | 1012 | 1225 |
| 84119 | 1 | $18,911 | 1 | 58.27 | 361 | 241 | 284 |
| 84115 | 1 | $19,797 | 1 | 63.24 | 698 | 492 | 609 |
| 84120 | 1 | $19,807 | 1 | 57.93 | 1159 | 865 | 1021 |
| 84118 | 1 | $19,949 | 1 | 64.09 | 453 | 357 | 409 |
| 84044 | 1 | $20,443 | 2 | 74.48 | 970 | 757 | 856 |
| 84128 | 2 | $20,502 | 1 | 59.98 | 74 | 51 | 71 |
| 84111 | 2 | $23,069 | 2 | 75 | 669 | 342 | 380 |
| 84123 | 2 | $23,296 | 2 | 74.76 | 976 | 713 | 829 |
| 84088 | 2 | $23,611 | 3 | 78.79 | 1678 | 1186 | 1487 |
| 84129 | 2 | $23,834 | 2 | 74.48 | 830 | 591 | 704 |
| 84084 | 2 | $24,260 | 2 | 77.02 | 1816 | 1191 | 1235 |
| 84081 | 2 | $24,554 | 3 | 77.91 | 1413 | 994 | 1288 |
| 84107 | 3 | $25,396 | 3 | 79.4 | 1508 | 917 | 1196 |
| 84102 | 3 | $26,141 | 3 | 80.45 | 313 | 183 | 226 |
| 84047 | 3 | $26,417 | 2 | 75.83 | 54 | 42 | 53 |
| 84096 | 3 | $28,722 | 5 | 90.66 | 1125 | 851 | 1149 |
| 84070 | 3 | $28,849 | 3 | 82.41 | 1378 | 864 | 1093 |
| 84065 | 3 | $29,558 | 5 | 92.55 | 1143 | 845 | 1270 |
| 84094 | 4 | $30,528 | 4 | 87.73 | 1056 | 633 | 843 |
| 84101 | 4 | $30,600 | 2 | 77.67 | 870 | 419 | 564 |
| 84106 | 4 | $31,440 | 3 | 83.93 | 890 | 450 | 754 |
| 84095 | 4 | $33,995 | 5 | 89.28 | 790 | 466 | 650 |
| 84124 | 4 | $35,433 | 4 | 89.19 | 1036 | 679 | 971 |
| 84109 | 4 | $36,024 | 5 | 89.74 | 1036 | 679 | 971 |
| 84103 | 4 | $36,325 | 4 | 85.41 | 580 | 335 | 403 |
| 84020 | 5 | $36,443 | 4 | 88.78 | 876 | 619 | 832 |
| 84093 | 5 | $37,033 | 5 | 90.73 | 1386 | 885 | 1154 |
| 84121 | 5 | $37,328 | 5 | 89.63 | 478 | 275 | 375 |
| 84117 | 5 | $38,282 | 4 | 87.65 | 1304 | 750 | 1121 |
| 84092 | 5 | $39,177 | 5 | 90.84 | 425 | 298 | 405 |
| 84105 | 5 | $39,472 | 4 | 88.76 | 794 | 477 | 698 |
| 84108 | 5 | $43,068 | 4 | 85.77 | 380 | 127 | 136 |

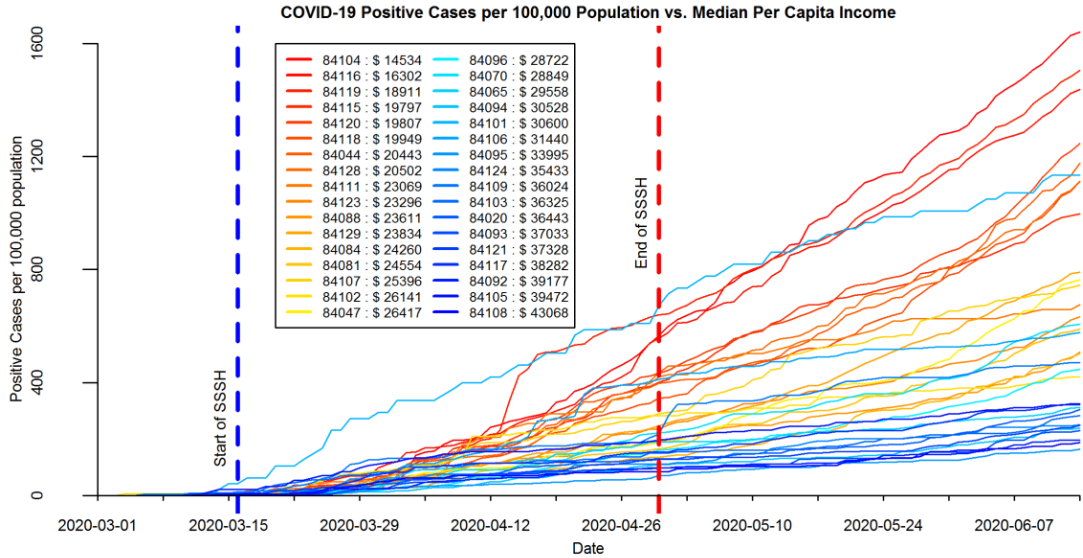

A

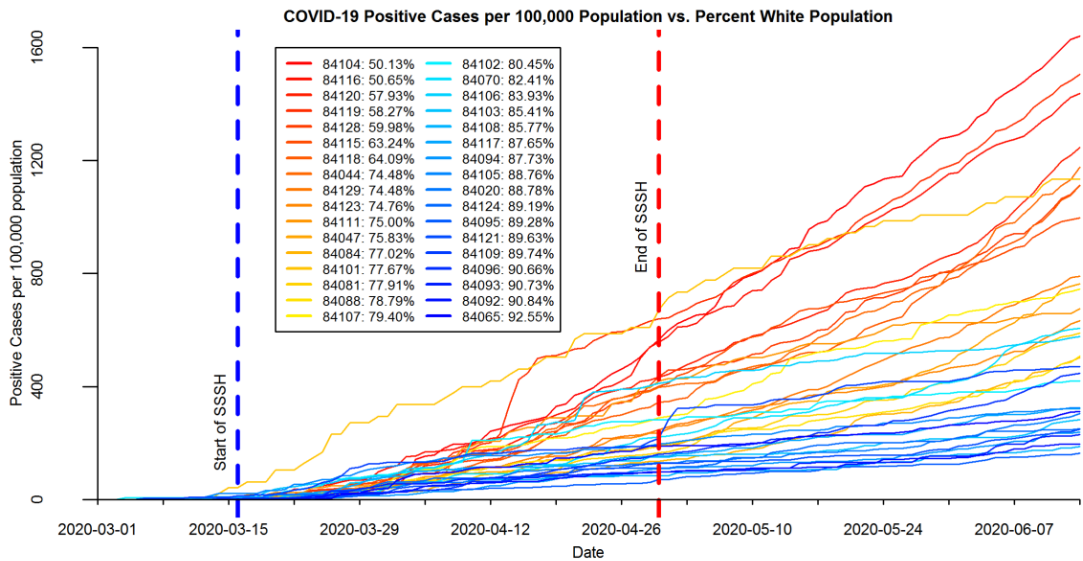

B

Figure S1. COVID-19 cases disaggregated by zip code and (A) per capita income and (B) percent white population. The dashed vertical lines show the start (blue) and end (red) of SSSH directives. The color scale ranges from red (lowest income/percent white population) to blue (highest income/percent white population).